\documentclass[graybox]{svmult}

\usepackage{helvet}         
\usepackage{courier}        
\usepackage{type1cm}        
\usepackage{makeidx}         
\usepackage{graphicx}        
\usepackage{multicol}        
\usepackage[bottom]{footmisc}

\newcommand\rf[1]{(\ref{eq:#1})}
\newcommand\lab[1]{\label{eq:#1}}
\newcommand\nonu{\nonumber}
\newcommand\br{\begin{eqnarray}}
\newcommand\er{\end{eqnarray}}
\newcommand\be{\begin{equation}}
\newcommand\ee{\end{equation}}

\newcommand\lb{\lbrack}
\newcommand\rb{\rbrack}
\newcommand\llangle{\left\langle}
\newcommand\rrangle{\right\rangle}

\newcommand\lcurl{\left\{}
\newcommand\rcurl{\right\}}
\renewcommand\({\left(}
\renewcommand\){\right)}

\newcommand\bc{\begin{center}}
\newcommand\ec{\end{center}}

\newcommand\partder[2]{\frac{{\partial {#1}}}{{\partial {#2}}}}

\renewcommand\a{\alpha}
\renewcommand\b{\beta}

\renewcommand\d{\delta}
\newcommand\eps{\epsilon}
\newcommand\vareps{\varepsilon}
\newcommand\g{\gamma}
\newcommand\G{\Gamma}

\newcommand\h{\frac{1}{2}}
\renewcommand\k{\kappa}
\renewcommand\l{\lambda}
\renewcommand\L{\Lambda}
\newcommand\m{\mu}
\newcommand\n{\nu}
\newcommand\om{\omega}

\newcommand\vp{\varphi}
\renewcommand\P{\Phi}
\newcommand\pa{\partial}

\newcommand\pr{\prime}

\newcommand\s{\sigma}

\renewcommand\t{\tau}

\newcommand\wti{\widetilde}

\newcommand\cA{{\mathcal A}}
\newcommand\cB{{\mathcal B}}
\newcommand\cC{{\mathcal C}}

\newcommand\cF{{\mathcal F}}

\newcommand\cH{{\mathcal H}}

\newcommand\cW{{\mathcal W}}

\newcommand{\ct}[1]{\cite{#1}}
\newcommand{\bib}[1]{\bibitem{#1}}

\newcommand\PRD[3]{\textsl{Phys. Rev.} \textbf{D#1} (#2) #3}

\newcommand\IJMPA[3]{\textsl{Int. J. Mod. Phys.} \textbf{A#1} (#2) #3}
\newcommand\IJMPD[3]{\textsl{Int. J. Mod. Phys.} \textbf{D#1} (#2) #3}

\newcommand\vpdot{\stackrel{.}{\varphi}}

\newcommand\adot{\stackrel{.}{a}}

\begin{document}

\title*{Wheeler-DeWitt Quantization of Gravity Models of Unified Dark 
Energy and Dark Matter}
\titlerunning{Wheeler-DeWitt Quantization of Unified Dark Energy and Dark Matter}
\author{Eduardo Guendelman, Emil Nissimov and Svetlana Pacheva}
\institute{Eduardo Guendelman \at Department of Physics, Ben-Gurion University of
the Negev, Beer-Sheva, Israel\\
\email{guendel@bgu.ac.il}
\and Emil Nissimov and Svetlana Pacheva
\at Institute for Nuclear Research and Nuclear Energy,
Bulgarian Academy of Sciences, Sofia, Bulgaria \\
\email{nissimov@inrne.bas.bg, svetlana@inrne.bas.bg}}
\maketitle

\abstract{
First, we describe the construction of a new type of gravity-matter models based 
on the formalism of non-Riemannian space-time volume forms - alternative 
generally covariant integration measure densities (volume elements) defined 
in terms of auxiliary antisymmetric tensor gauge fields. Here gravity couples 
in a non-conventional way to two distinct scalar fields providing a unified 
Lagrangian action principle description of: (i) the evolution of both ``early''
and ``late'' Universe - by the ``inflaton'' scalar field; (ii) dark energy and
dark matter as a unified manifestation of a single material entity - 
the ``darkon'' scalar field. A physically very interesting phenomenon occurs 
when including in addition interactions with the electro-weak model 
bosonic sector - we obtain a gravity-assisted dynamical generation of 
electro-weak spontaneous gauge symmetry breaking in the post-inflationary 
``late'' Universe, while the Higgs-like scalar remains massless in the 
``early'' Universe.
Next, we proceed to the Wheeler-DeWitt minisuperspace quantization of the 
above models. The ``darkon'' field plays here the role of cosmological ``time''. 
In particular, we show the absence of cosmological space-time singularities.
}

\section{Introduction}
\label{intro}

Among the most important paradigms at the interface of particle physics and 
cosmology \ct{general-cit-1}-\ct{general-cit-7} one should mention:

\begin{itemize}
\item
(i) The nature of dark energy and dark matter -- both ``dark'' species
occupying around 70\% and 25\% of the matter content of the ``late'' (today's) 
Universe, respectively, continue to be the two most unexplained ``mysteries'' 
in cosmology and astrophysics (for a background, see 
\ct{dark-energy-observ-1}-\ct{DM-rev-3}).
\item
(ii) The interplay between the cosmological dynamics and the evolution of the
symmetry breaking patterns along the history of the Universe -- specifically,
for the present epoch's phase of slowly accelerating Universe (dark energy
domination) see \ct{dark-energy-observ-1}-\ct{dark-energy-observ-7}, and for 
a recent general account see \ct{rubakov,calcagni}.
\end{itemize}

There exist a multitude of proposals for an adequate description of dark energy's 
and dark matter's dynamics within the framework of standard general relativity 
or its modern extensions, among them: ``Chaplygin gas'' models 
\ct{chaplygin-1,chaplygin-2,chaplygin-3}, ``purely kinetic k-essence'' models 
\ct{purely-kinetic-k-essence-1,purely-kinetic-k-essence-2}, 
``mimetic'' dark matter models \ct{mimetic-1}-\ct{mimetic-4}.

Addressing issue (i) above, in Section 2  we will briefly review our own approach
\ct{dusty-1,dusty-2} (for some earlier works, see also 
\ct{eduardo-singleton,eduardo-ansoldi})
to one of the principal challenge in modern cosmology to understand
theoretically from first principles the nature of both ``dark'' species as a
manifestation of the dynamics of a single entity of matter. In the simplest
setting we achieve unified description of dark energy and dark matter based on a 
class of generalized non-canonical models of gravity interacting with a single
scalar {\em ``darkon''} field employing the method of non-Riemannian
volume-forms on the pertinent spacetime manifold,
\textsl{i.e.}, non-Riemannian volume elements.
Originally \ct{TMT-orig-1,TMT-orig-2,TMT-orig-3} this approach was proposed 
as introducing alternative generally covariant integration measure densities 
in terms of auxiliary ``measure'' scalar fields. 
Later \ct{susyssb-1,belgrade-14,grav-bags} it was reformulated in a more consistent
geometrical setting, namely, the non-Riemannian volume-forms are constructed in 
terms of auxiliary higher-rank antisymmetric tensor gauge fields, which were
shown to be essentially pure-gauge degrees of freedom, \textsl{i.e.}, {\em no}
additional propagating field-theoretic (gravitational) degrees of freedom are 
introduced.

Next, addressing issue (ii) we extend \ct{emergent,grf-essay} the above  
non-canonical gravity-matter model by adding coupling to a second scalar 
{\em ``inflaton''} field describing the universe's evolution in a 
unified way (``quintessence''), as well as coupling to the fields of the 
electroweak bosonic sector. In this way we obtain a {\em gravity-assisted} 
generation of electro-weak spontaneous gauge symmetry breaking in the 
post-inflationary ``late'' Universe, while the Higgs-like scalar remains 
massless in the ``early'' Universe \ct{grf-essay,BJP-3rd-congress}.

In Section 3 we perform  Wheeler-DeWitt \ct{WDW-1,WDW-2} minisuperspace
quantization of the above models. The ``darkon'' field plays the role of 
cosmological ``time'' in the pertinent Wheeler-De Witt equation in the 
``early'' universe. We show explicitly the absence of cosmological 
singularities in the wave function of the universe.

\section{Quintessence, Unified Dark Energy and Dark Matter, and Gravity-Assisted 
Higgs Mechanism}
\label{quintess-DE-DM-Higgs}

\subsection{Hidden Noether Symmetry and Unification of Dark Energy and 
Dark Matter}
\label{hidden}

First, let us consider the following simple particular case of a non-conventional 
gravity-scalar-field action -- a member of the general class of the 
non-Riemannian-volume-element-based gravity-matter theories \ct{emergent,grav-bags}
(for simplicity we use units with the Newton constant $G_N = 1/16\pi$):
\be
S = \int d^4 x \sqrt{-g}\, R +
\int d^4 x \bigl(\sqrt{-g}+\P(C)\bigr) L(u,Y) \; .
\lab{TMT-0}
\ee
Here $R$ denotes the standard Riemannian scalar curvature for the pertinent
Riemannian metric $g_{\m\n}$. 
In the second term in \rf{TMT-0} -- the scalar field Lagrangian is coupled
{\em symmetrically} to two mutually independent spacetime volume-elements 
-- the standard Riemannian $\sqrt{-g}$ and to an alternative 
non-Riemannian one:
\be
\P(C) = \frac{1}{3!}\vareps^{\m\n\k\l} \pa_\m C_{\n\k\l} \; .
\lab{mod-measure}
\ee

$L(u,Y)$ is general-coordinate invariant Lagrangian of a single scalar field 
$u (x)$, the simplest example being:
\be
L(u,Y) = Y - V(u) \quad ,\quad Y \equiv - \h g^{\m\n}\pa_\m u \pa_\n u \; ,
\lab{standard-L}
\ee
Crucial new property -- we obtain {\em dynamical constraint} on $L(u,Y)$ 
as a result of the equations of motion w.r.t. 
$C_{\m\n\l}$:
\be
\pa_\m L (u,Y) = 0 \; \longrightarrow \; L(u,Y) = - 2M_0 = {\rm const} \; ,
\lab{L-const}
\ee
\textsl{i.e.}, $Y = V(u) - 2M_0$.
$M_0$ will play the role of dynamically generated cosmological constant.

A second crucial property -- {\em hidden strongly nonlinear Noether symmetry}
of scalar field action in \rf{TMT-0} -- is due to the presence of 
the non-Riemannian volume element $\P (C)$.
The scalar field action is invariant (up to a total derivative) under the
following nonlinear symmetry transformations:
\br
\d_\eps u = \eps \sqrt{Y} \quad ,\quad \d_\eps g_{\m\n} = 0 \quad ,\quad
\d_\eps \cC^\m = - \eps \frac{1}{2\sqrt{Y}} g^{\m\n}\pa_\n u 
\bigl(\P(C) + \sqrt{-g}\bigr)  \; ,
\lab{hidden-sym}
\er
where $\cC^\m \equiv \frac{1}{3!} \vareps^{\m\n\k\l} C_{\n\k\l}$.

Then, standard Noether procedure yields a conserved current:
\be
\nabla_\m J^\m = 0 \quad ,\quad
J^\m \equiv - \Bigl(1+\frac{\P(C)}{\sqrt{-g}}\Bigr)\sqrt{2Y} 
g^{\m\n}\pa_\n u 
\lab{J-conserv}
\ee
The energy-momentum tensor $T_{\m\n}$ and $J^\m$ \rf{J-conserv} can be cast into
a relativistic hydrodynamical form (taking into account \rf{L-const}):
\be
T_{\m\n} = - 2M_0 g_{\m\n} + \rho_0 u_\m u_\n \quad ,\quad J^\m = \rho_0 u^\m \; ,
\lab{T-J-hydro}
\ee
where the pressure $p = - 2M_0 = {\rm const}$ and:
\be
\rho_0 \equiv \Bigl(1+\frac{\P(C)}{\sqrt{-g}}\Bigr)\, 2Y 
\;\;\; ,\;\; 
u_\m \equiv - \frac{\pa_\m u}{\sqrt{2Y}} \;\; ,\;\; 
 u^\m u_\m = -1 \; .
\lab{rho-0-def}
\ee 
The total energy density is
~$\rho = \rho_0 - p = 2M_0 + \Bigl(1+\frac{\P(C)}{\sqrt{-g}}\Bigr)\, 2Y$.

Because of the constant pressure ($p=-2M_0$) $\nabla^\n T_{\m\n}=0$ implies
{\em both} hidden Noether symmetry current $J^\m = \rho_0 u^\m$ conservation, 
as well as {\em geodesic fluid motion}:
\be
\nabla_\m \bigl(\rho_0 u^\m\bigr) = 0 \quad ,\quad 
u_\n \nabla^\n u_\m = 0 \; .
\lab{dust-geo}
\ee

Therefore, $T_{\m\n} = - 2M_0 g_{\m\n} + \rho_0 u_\m u_\n$ 
represents an exact sum of two contributions of the two dark species:
\br
p = p_{\rm DE} + p_{\rm DM} \quad,\quad \rho = \rho_{\rm DE} + \rho_{\rm DM}
\lab{DE+DM-1} \\
p_{\rm DE} = -2M_0\;\; ,\;\; \rho_{\rm DE} = 2M_0 \quad ; \quad
p_{\rm DM} = 0\;\; ,\;\; \rho_{\rm DM} = \rho_0 \; ,
\lab{DE+DM-2}
\er
\textsl{i.e.}, the dark matter component is a dust fluid flowing along
geodesics. This is explicit unification of dark energy and dark matter
originating from the dynamics of a single scalar field - the ``darkon'' $u$.

\subsection{Quintessential Inflation and Unified Dark Energy and Dark Matter}
\label{quintess-DE-DM}

We will now extend our previous gravity-``darkon'' model to gravity coupled to 
both ``inflaton'' $\vp (x)$ and ``darkon'' $u(x)$ scalar fields within the
non-Riemannian volume-form formalism, as well as we will also add 
coupling to the bosonic sector of the electro-weak model:
\br
S = \int d^4 x\,\P (A) \Bigl\lb g^{\m\n} R_{\m\n}(\G) + L_1 (\vp,X)
+ L_2 (\s,\nabla\s;\vp) \Bigr\rb +
\nonu \\
\int d^4 x\,\P (B) \Bigl\lb U (\vp) + L_3 (\cA,\cB)
+ \frac{\P (H)}{\sqrt{-g}}\Bigr\rb
+ \int d^4 x \bigl(\sqrt{-g}+\P(C)\bigr) L(u,Y) \; .
\lab{TMMT-1}
\er
Here the following notations are used:
\begin{itemize}
\item
$\P(A)=\frac{1}{3!}\vareps^{\m\n\k\l} \pa_\m A_{\n\k\l}$~ and 
~$\P(B)= \frac{1}{3!}\vareps^{\m\n\k\l} \pa_\m B_{\n\k\l}$~ -- ~two new 
independent non-Riemannian volume-forms (non-Riemannian volume elements)
apart from $\P(C)$;
\end{itemize}

\begin{itemize}
\item
$\P (H) = \frac{1}{3!}\vareps^{\m\n\k\l} \pa_\m H_{\n\k\l}$  
is the dual field-strength of an additional auxiliary tensor gauge field 
$H_{\n\k\l}$ crucial for the consistency of \rf{TMMT-1}.
\item
Important -- we use Palatini formalism:  
$R=g^{\m\n} R_{\m\n}(\G)$ ; $g_{\m\n}$, $\G^\l_{\m\n}$ -- metric and affine
connection are {\em apriori} independent.
\item
$\s \equiv (\s_a)$ is a complex $SU(2)\times U(1)$ iso-doublet Higgs-like scalar field
with a Lagrangian:
\be
L_2 (\s,\nabla\s;\vp) = - g^{\m\n} \bigl(\nabla_\m \s_a)^{*}\nabla_\n \s_a 
- V_0 (\s)e^{\a\vp} \; .
\lab{L-sigma}
\ee
The gauge-covariant derivative acting on $\s$ reads:
\be
\nabla_\m \s = 
\Bigl(\pa_\m - \frac{i}{2} \t_A \cA_\m^A - \frac{i}{2} \cB_\m \Bigr)\s \; ,
\lab{cov-der}
\ee
with $\h \t_A$ ($\t_A$ -- Pauli matrices, $A=1,2,3$) indicating the $SU(2)$ 
generators. 
\end{itemize}

\begin{itemize}
\item
The ``bare'' $\s$-field potential is of the same form as the standard Higgs
potential: 
\be
V_0 (\s) = \frac{\l}{4} \((\s_a)^{*}\s_a - \m^2\)^2 \; .
\lab{standard-higgs}
\ee
\item
The $SU(2)\times U(1)$ gauge field action $L(\cA,\cB)$ is of the standard Yang-Mills form 
(all $SU(2)$ indices $A,B,C = (1,2,3)$):
\br
L_3(\cA,\cB) = - \frac{1}{4g^2} F^2(\cA) - \frac{1}{4g^{\pr\,2}} F^2(\cB)
\; , \phantom{aaaaa}
\lab{EW-gauge-L} \\
F^2(\cA) \equiv F^A_{\m\n} (\cA) F^A_{\k\l} (\cA) g^{\m\k} g^{\n\l} \;\; ,\;\;
F^2(\cB) \equiv F_{\m\n} (\cB) F_{\k\l} (\cB) g^{\m\k} g^{\n\l} \; ,
\nonu \\
F^A_{\m\n} (\cA) = 
\pa_\m \cA^A_\n - \pa_\n \cA^A_\m + \eps^{ABC} \cA^B_\m \cA^C_\n \;\; ,\;\;
F_{\m\n} (\cB) = \pa_\m \cB_\n - \pa_\n \cB_\m \; .
\nonu
\er
$\cA_\m^A$ ($A=1,2,3$) and $\cB_\m$ denote the corresponding $SU(2)$ and $U(1)$ 
electroweak gauge fields.
\end{itemize}

\begin{itemize}
\item
The ``inflaton'' $\vp$ Lagrangian terms are given by:
\br
L_1 (\vp,X) = X - V_1(\vp) \quad, \quad
X \equiv - \h g^{\m\n} \pa_\m \vp \pa_\n \vp \; ,
\lab{L-1} \\
V_1(\vp) = f_1 \exp \{\a\vp\} \quad ,\quad U(\vp) \equiv f_2 \exp \{2\a\vp\} \; ,
\lab{L-2}
\er
where $\a, f_1, f_2$ are dimensionful positive parameters. 
\item
The form of the action \rf{TMMT-1} is fixed by the requirement of invariance
under global Weyl-scale transformations:
\br
g_{\m\n} \to \l g_{\m\n} \;,\; \G^\m_{\n\l} \to \G^\m_{\n\l} \; ,\; 
\vp \to \vp - \frac{1}{\a}\ln \l\; ,
\nonu \\
A_{\m\n\k} \to \l A_{\m\n\k} \; ,\; B_{\m\n\k} \to \l^2 B_{\m\n\k} \; ,\; 
H_{\m\n\k} \to H_{\m\n\k} \; ,
\lab{scale-transf}
\er
and the electro-weak sector $(\s,\cA,\cB)$ is inert w.r.t. \rf{scale-transf}.
\end{itemize}

Eqs. of motion w.r.t. affine connection $\G^\m_{\n\l}$ yield a solution for
the latter as a Levi-Civita connection:
\be
\G^\m_{\n\l} = \G^\m_{\n\l}({\bar g}) = 
\h {\bar g}^{\m\k}\(\pa_\n {\bar g}_{\l\k} + \pa_\l {\bar g}_{\n\k} 
- \pa_\k {\bar g}_{\n\l}\) \; ,
\lab{G-eq}
\ee
w.r.t. to the {\em Weyl-rescaled metric} ${\bar g}_{\m\n}$:
\be
{\bar g}_{\m\n} = \chi_1 g_{\m\n} \quad ,\quad
\chi_1 \equiv \frac{\P_1 (A)}{\sqrt{-g}} \; . 
\lab{bar-g}
\ee
Transition from original metric $g_{\m\n}$ to ${\bar g}_{\m\n}$:
{\em ``Einstein-frame''}, where the gravity eqs. of motion are written in
the standard form of Einstein's equations:
$R_{\m\n}({\bar g}) - \h {\bar g}_{\m\n} R({\bar g}) = \h T^{\rm eff}_{\m\n}$ 
with an appropriate {\em effective} energy-momentum tensor given in terms
of an Einstein-frame matter Lagrangian $L_{\rm eff}$ (see \rf{L-eff-total} below).

Solutions of the eqs. of motion of the action \rf{TMMT-1} w.r.t. auxiliary 
tensor gauge fields $A_{\m\n\l}$, $B_{\m\n\l}$ and $H_{\m\n\l}$ yield:
\br
\frac{\P(B)}{\sqrt{-g}} \equiv \chi_2 = {\rm const} \quad ,\quad
R + L_1 (\vp,X) + L_2 (\s,\nabla\s;\vp)= M_1 = {\rm const} \; ,
\nonu \\
U (\vp) + L_3 (\cA,\cB) + \frac{\P (H)}{\sqrt{-g}} = - M_2  = {\rm const} \; .
\lab{integr-const}
\er
Here $M_1$ and $M_2$ are arbitrary dimensionful and $\chi_2$
arbitrary dimensionless integration constants, similar to $M_0$ \rf{L-const}.

Within the canonical Hamilton formalism we have shown 
\ct{belgrade-14,susyssb-2,grav-bags} that
$M_0,\,M_{1,2},\,\chi_2$ are the only remnant of the auxiliary gauge fields
$C_{\m\n\l},\,A_{\m\n\l},$ $\,B_{\m\n\l},\,H_{\m\n\l}$ entering \rf{TMMT-1} -- 
they have the meaning of conserved Dirac-constrained canonical
momenta conjugated to some of the components of the latter.




We derive from \rf{TMMT-1} the physical {\em Einstein-frame} theory w.r.t. 
Weyl-rescaled Einstein-frame metric ${\bar g}_{\m\n}$ \rf{bar-g} and perform an
additional ``darkon'' field redefinition $u \to {\wti u}$: 
\be
\;\; \partder{{\wti u}}{u}=\bigl( V_1(u) - 2M_0\bigr)^{-\h} \quad ; \quad
Y \to {\wti Y} = - \h {\bar g}^{\m\n}\pa_\m {\wti u} \pa_\n {\wti u} \; .
\lab{darkon-redef}
\ee
The Einstein-frame action reads:
\be
S = \int d^4 x \sqrt{-{\bar g}} \Bigl\lb R({\bar g}) + 
L_{\rm eff}\bigl(\vp, {\bar X}, {\wti Y};\s,{\bar X}_\s,\cA,\cB\bigr)\Bigr\rb \; ,
\lab{einstein-frame-1}
\ee
where (now the kinetic terms are given in terms of the Einstein-frame metric
\rf{bar-g}, 
\textsl{e.g.} ${\bar X} = - \h {\bar g}^{\m\n}\pa_\m \vp \pa_\n \vp$, etc.):
\br
L_{\rm eff}\bigl(\vp, {\bar X}, {\wti Y};\s,{\bar X}_\s,\cA,\cB\bigr) = {\bar X} 
- {\wti Y}\Bigl(V_1(\vp) + V_0 (\s)e^{\a\vp} + M_1\Bigr)
\nonu \\
+ {\wti Y}^2 \Bigl\lb \chi_2 (U(\vp) + M_2) - 2 M_0\Bigr\rb + L\lb\s,{\bar X}_\s,\cA,\cB\rb 
\; ,
\phantom{aaaa}
\lab{L-eff-total} 
\er
with $L\lb\s,{\bar X}_\s,\cA,\cB\rb \equiv
- {\bar g}^{\m\n} \bigl(\nabla_\m \s_a)^{*}\nabla_\n \s_a
- \frac{\chi_2}{4g^2} {\bar F}^2(\cA) - \frac{\chi_2}{4g^{\pr\,2}} {\bar F}^2(\cB)$.

For static (spacetime idependent) scalar field configurations we obtain from 
\rf{L-eff-total} the following Einstein-frame effective scalar
``inflaton+Higgs'' effective potential:
\br
U_{\rm eff}\bigl(\vp,\s\bigr) = 
\frac{\Bigl(V_1 (\vp) + V_0 (\s)e^{\a\vp} + M_1\Bigr)^2}{4\bigl\lb \chi_2 (U(\vp) +
M_2) - 2 M_0\bigr\rb}
\nonu \\
= \frac{\Bigl\lb \Bigl(f_1 + \frac{\l}{4} \((\s_a)^{*}\s_a -
\m^2\)^2\Bigr) e^{\a\vp}
+ M_1\Bigr\rb^2}{4\bigl\lb \chi_2 (f_2 e^{2\a\vp} + M_2) - 2 M_0\bigr\rb} \; .
\lab{U-eff-total} 
\er

$U_{\rm eff}\bigl(\vp,\s\bigr)$ has few remarkable properties.
First, $U_{\rm eff}\bigl(\vp,\s\bigr)$ possesses two infinitely large flat
regions as function of $\vp$ (when $\s$ is fixed):\\
$\phantom{aaaa}$ 
(a) $(-)$ flat region for large negative values of the ``inflaton'' $\vp$;\\
$\phantom{aaaa}$
(b) $(+)$ flat region and large positive values of $\vp$,\\
respectively, as depicted in Fig.1..

\begin{figure}
\begin{center}
\includegraphics[width=9cm,keepaspectratio=true]{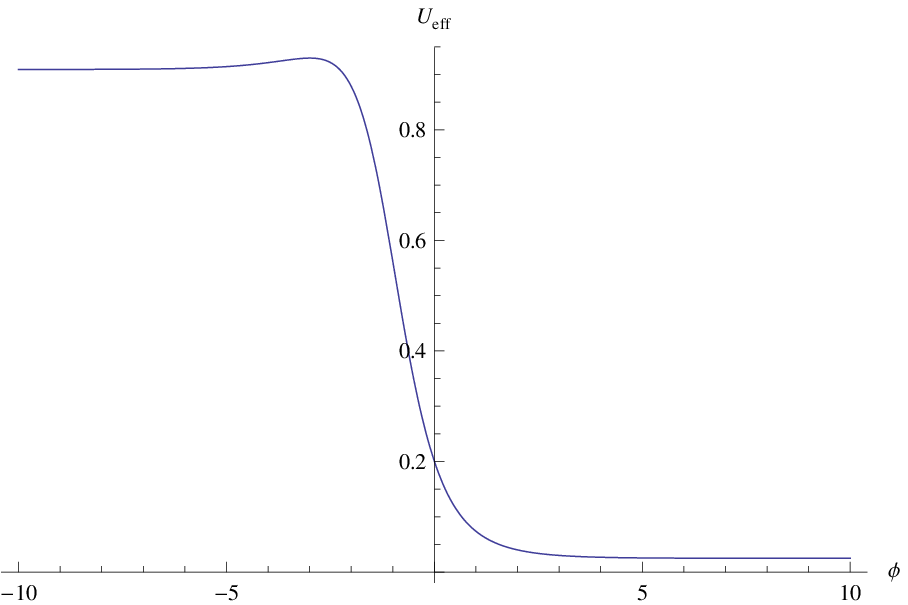}
\caption{Qualitative shape of the effective scalar potential $U_{\rm eff}$ 
\rf{U-eff-total} as function of $\vp$ at $\s = {\rm fixed}$ for $M_1 > 0$.}
\end{center}
\end{figure}

\begin{itemize}
\item 
In the $(+)$ flat region (large positive ``inflaton'' values) \rf{U-eff-total} 
reduces to:
\be
U_{\rm eff}\bigl(\vp,\s\bigr) \simeq U_{(+)}(\s) =
\frac{\Bigl(\frac{\l}{4} \((\s_a)^{*}\s_a - \m^2\)^2 + f_1\Bigr)^2}{4\chi_2 f_2} \; .
\lab{U-plus-higgs}
\ee
\item
\rf{U-plus-higgs} yields as a lowest lying vacuum the Higgs one:
\be
|\s| = \m  \; ,
\lab{higgs-vac}
\ee
\textsl{i.e.}, 
we obtain the
standard spontaneous breakdown of $SU(2)\times U(1)$ ~gauge symmetry.
\item
At the Higgs vacuum \rf{higgs-vac} we get from \rf{U-plus-higgs}
a dynamically generated cosmological constant $\L_{(+)}$: 
\be
U_{(+)}(\m) \equiv 2\L_{(+)} = \frac{f_1^2}{4\chi_2 f_2} \; .
\lab{CC-eff-plus}
\ee
\end{itemize}

\begin{itemize}
\item
If we identify the integration constants in \rf{U-eff-total} with the fundamental 
constants of Nature -- $M_{Pl}$ (Planck mass) and $M_{EW}$ (electro-weak mass scale)
as $f_1 \sim M^4_{EW}$, $f_2 \sim M^4_{Pl}$, 
we are then naturally led to a very small vacuum energy density:
\be
U_{(+)}(\m) \sim M^8_{EW}/M^4_{Pl} \sim 10^{-122} M^4_{Pl} \; ,
\lab{U-plus-magnitude}
\ee
which is the right order of magnitude for the present epoch's vacuum energy density
according to \ct{arkani-hamed}.
Therefore, we can identify the $(+)$ flat region (large positive ``inflaton'' values)
of $U_{\rm eff}$ \rf{U-eff-total} as describing the present ``late'' universe.
\end{itemize}

\begin{itemize}
\item
In the $(-)$ flat region (large negative ``inflaton'' values) \rf{U-eff-total} 
reduces to:
\be
U_{\rm eff}\bigl(\vp,\s\bigr) \simeq U_{(-)} \equiv 
\frac{M_1^2}{4(\chi_2\,M_2-2M_0)} \; .
\lab{U-minus} 
\ee
If we take 
$M_1 \sim M_2 \sim 10^{-8} M_{Pl}^4$ and $M_0 \sim M_{EW}^4$, then 
the vacuum energy density $ U_{(-)}$ \rf{U-minus} becomes
~$U_{(-)} \sim 10^{-8} M_{Pl}^4$,
which conforms to the Planck Collaboration data \ct{Planck-1,Planck-2} 
for the energy scale of inflation (of order $10^{-2} M_{Pl}$). This allows
to identify the $(-)$ flat region (large negative ``inflaton'' values) of
the ``inflaton+Higgs'' effective potential \rf{U-eff-total} as describing the
``early'' universe, in particular, the inflationary epoch.
\item
In the $(-)$ flat region the effective potential \rf{U-minus} is $\s$-field idependent.
Thus, the Higgs-like iso-doublet scalar field $\s_a$ remains {\em massless} in 
the ``early'' (inflationary) Universe and accordingly there is 
{\em no electro-weak spontaneous symmetry breaking} there.
\end{itemize}

\section{Wheeler-De Witt Minisuperspace Quantization}
\label{WDW}

For simplicity here we will consider the unified dark energy/dark matter
``quintessential'' model \rf{TMMT-1} without the coupling to the bosonic 
electro-weak sector. The corresponding Einstein-frame action reads:
\be
S = \int d^4 x \sqrt{-{\bar g}} \Bigl\lb R({\bar g}) + 
L_{\rm eff}\bigl(\vp, {\bar X}, {\wti Y}\bigr)\Bigr\rb \; ,
\lab{einstein-frame}
\ee
where (recall ${\bar X} = - \h {\bar g}^{\m\n}\pa_\m \vp \pa_\n \vp$~
and ~${\wti Y} = - \h {\bar g}^{\m\n}\pa_\m {\wti u} \pa_\n {\wti u}$):
\be
L_{\rm eff}\bigl(\vp, {\bar X}, {\wti Y}\bigr) = {\bar X} 
- {\wti Y}\Bigl(V(\vp) - M_1\Bigr) +
{\wti Y}^2 \Bigl\lb \chi_2 (U(\vp) + M_2) - 2 M_0\Bigr\rb \; ,
\lab{L-eff-total-0}
\ee
To study cosmological implications of \rf{einstein-frame} we perform a 
Friedmann-Lemaitre-Robertson-Walker ({\em FLRW}) reduction to the class of 
FLRW metrics:
\be
ds^2 = {\bar g}_{\m\n} dx^\m dx^\n = - N^2(t) dt^2 + a^2(t) d{\vec x}.d{\vec x}
\lab{FLRW}
\ee
and take the ``inflaton'' and ``darkon'' to be time-dependent only, 
\textsl{i.e.}:
\be
{\bar X} = \h \vpdot^2 \quad,\quad {\wti Y} = \h w^2 \;\;, \;\;
w \equiv \frac{d{\wti u}}{dt} \; .
\lab{X-Y-FLRW}
\ee

The FLRW reduced action corresponding to \rf{einstein-frame} reads:
\br
S_{\rm FLRW} = \int dt \lcurl -\frac{1}{N} 6 a\adot^2 
+ N a^3 \Bigl\lb \frac{\vpdot^2}{2N^2} 
- (f_1 e^{\a\vp} + M_1) \frac{w^2}{2N^2}\right. \\
\left. + \Bigl(\chi_2 (f_2 e^{2\a\vp} + M_2) 
- 2M_0\Bigr) \frac{w^4}{4N^4}\Bigr\rb\rcurl
\lab{FRLW-action}
\er

Calculating the canonically conjugated momenta $p_a, p_\vp, p_{\wti u}$, we
arrive at the canonical FLRW Hamiltonian:
\br
\cH = N \cH_{WDW} = 
N \lcurl - \frac{p_a^2}{24a} + \frac{p_\vp^2}{a^3} + p_{\wti u} w \right. \\
\left. + a^3 \Bigl\lb (f_1 e^{\a\vp} + M_1) \frac{w^2}{2} 
- \Bigl(\chi_2 (f_2 e^{2\a\vp} + M_2) - 2M_0\Bigr)\frac{w^4}{4}\Bigr\rb\rcurl
\lab{can-Ham}
\er
$\cH$ turns out to be pure first-class constraint $\cH_{WDW}$ a'la Dirac with the 
lapse $N$ as Lagrange multiplier.

In \rf{can-Ham} the ``darkon'' velocity $w$ is determined as function of the
canonical variables $(a,\vp,p_{\wti u})$ being the real root (for all values of 
$(a,\vp,p_{\wti u})$) of the cubic algebraic equation:
\be
w^3 - 3 A(\vp)\, w - 2\frac{B(\vp,p_{\wti u})}{a^3} =0 \,
\lab{cubic-eq}
\ee
where the coefficients are given by:
\br
A(\vp) \equiv 
\frac{1}{3}\,\frac{(f_1 e^{\a\vp} + M_1)}{\chi_2 (f_2 e^{2\a\vp} + M_2) - 2M_0}
\;, 
\nonu \\
B(\vp,p_{\wti u}) \equiv 
\frac{p_{\wti u}}{2}\,\frac{1}{\chi_2 (f_2 e^{2\a\vp} + M_2) - 2M_0}\; .
\lab{def-A-B}
\er
The solution of \rf{cubic-eq} for $w=w(a,\vp,p_{\wti u})$ reads:
\be
w=\mathrm{sign}(B(\vp,p_{\wti u})) |A(\vp)|^{1/2} |\xi|^{-1/6}
\Bigl\lb \bigl(1+\sqrt{1-\xi}\bigr)^{1/3} + \bigl(1-\sqrt{1-\xi}\bigr)^{1/3}\Bigr\rb
\lab{w-sol}
\ee
where $\xi \equiv \xi(a,\vp,p_{\wti u}) 
= \frac{A^3(\vp)}{9 B^2(\vp,p_{\wti u})}\, a^6$.

Quantization of the Dirac-constrained canonical Hamilton \rf{can-Ham} 
yields the Wheeler-De Witt (WDW) equation for the wave
function of the universe $\Psi = \Psi (a,\vp; p_{\wti u})$:
\be
{\widehat \cH}_{WDW} \Psi (a,\vp; p_{\wti u}) = 0 \; ,
\lab{WDW-eq}
\ee
where ${\widehat \cH}_{WDW}$ is the quantum version of $\cH_{WDW}$ in \rf{can-Ham}.
We resolve the ordering ambiguity there by changing variables:
\be
a \to {\wti a} = \frac{4}{\sqrt{3}} a^{3/2} \; ,
\lab{a-tilde}
\ee
and taking the special operator ordering:
\be
\frac{p_a^2}{24 a} \to 
\h\frac{1}{\sqrt{12 a}}{\widehat p}_a \frac{1}{\sqrt{12 a}}{\widehat p}_a
= -\h \frac{\pa^2}{\pa {\wti a}^2} \; .
\lab{operator-order}
\ee
The WDW operator ${\widehat \cH}_{WDW}$ becomes:
\be
{\widehat \cH}_{WDW} = \h \frac{\pa^2}{\pa {\wti a}^2} +
\frac{8}{3{\wti a}^2} {\widehat p}^2_\vp + \frac{3}{4} p_{\wti u} w +
\frac{3}{64} w^2 {\wti a}^2 (f_1 e^{\a\vp} + M_1) \; ,
\lab{WDW-operator}
\ee
where ${\widehat p}_\vp = -i \pa/\pa \vp$ and $w=w({\wti a},\vp,p_{\wti u})$ is
the solution \rf{w-sol} of the cubic equation \rf{cubic-eq}.

The final form of WDW equation reads:
\br
\Bigl\lb \h \Bigl(\partder{}{{\wti a}}\Bigr)^2 +
\frac{8}{3{\wti a}^2} {\widehat p}^2_\vp  + U({\wti a},\vp,p_{\wti u})\Bigr\rb
\Psi ({\wti a},\vp; p_{\wti u}) = 0 \; ,
\lab{WDW-eq-final} \\
U({\wti a},\vp,p_{\wti u}) \equiv 
\frac{{\wti a}^2 (f_1 e^{\a\vp} + M_1)^2}{
64(\chi_2 f_2 e^{2\a\vp} + \chi_2 M_2 - 2 M_0)} 
\cF \bigl(\xi({\wti a},\vp,p_{\wti u})\bigr)) \;
\lab{U-def}
\er
with the following notations:
\br
\xi({\wti a},\vp,p_{\wti u}) \equiv 
\frac{{\wti a}^4 (f_1 e^{\a\vp} + M_1)^3}{
192 p^2_{\wti u}(\chi_2 f_2 e^{2\a\vp} + \chi_2 M_2 - 2 M_0)} \; ,
\lab{xi-def} \\
\cF (\xi) \equiv \xi^{-1/3} 
\Bigl\lb \bigl(1+\sqrt{1-\xi}\bigr)^{1/3} + \bigl(1-\sqrt{1-\xi}\bigr)^{1/3}\Bigr\rb
\nonu \\
\times \Bigl\lb 2 \xi^{-1/3} + 
\bigl(1+\sqrt{1-\xi}\bigr)^{1/3} + \bigl(1-\sqrt{1-\xi}\bigr)^{1/3}\Bigr\rb \; .
\lab{cF-def}
\er
Analytic solutions of \rf{WDW-eq-final} can be found when the ``inflaton'' $\vp$
is either on the $(-)$ flat region ($\vp$ large negative -- ``early'' universe) 
or on $(+)$ flat region ($\vp$ large positive -- ``late''/nowadays universe),
cf. Fig.1 above.

In the $(+)$ flat region of the ``inflaton'' $\vp$ (``late'' universe) the
WDW eq.\rf{WDW-eq-final} reduces to the quantum mechanical Schr{\"o}dinger
equation:
\br
\Bigl\lb \h \frac{\pa^2}{\pa {\wti a}^2} + \cW_{(+)} ({\wti a},p_\vp)\Bigr\rb
\Psi ({\wti a},p_\vp) = 0 \; ,
\lab{WDW-plus} \\
\cW_{(+)} ({\wti a},p_\vp) \equiv \frac{3 f_1^2}{64\chi_2 f_2}\, {\wti a}^2
+ \frac{8 p^2_\vp}{3}\, {\wti a}^{-2} \quad, \;\; p_\vp - \mathrm{small}\; .
\lab{W-plus} 
\er
The solution of \rf{WDW-plus} reads (here $c_{1,2}$ are constants):
\br
\Psi ({\wti a},p_\vp) = \sqrt{{\wti a}}\Bigl\lb c_1
J_{\frac{1}{4}\sqrt{1-\g}} \bigl(\h\b{\wti a}^2\bigr) +
c_2 J_{-\frac{1}{4}\sqrt{1-\g}} \bigl(\h\b{\wti a}^2\bigr)\Bigr\rb \; ,
\lab{WDW-sol-plus} \\
\b \equiv \sqrt{\frac{3 f_1^2}{32\chi_2 f_2}} \quad ,\quad
\g \equiv \frac{64}{3}p^2_\vp \;\;\; (\g - \mathrm{small}) \; ,
\lab{WDW-sol-plus-defs} \\
\Psi ({\wti a},p_\vp) \simeq \mathrm{const}\;\, {\wti a}^{\h (1-\sqrt{1-\g})} 
\;\; \mathrm{for} \;\; {\wti a} \to 0 \; ,
\lab{no-singularity-plus}
\er
\textsl{i.e.}, the wave function \rf{WDW-sol-plus} vanishes at ${\wti a}=0$.

Similarly, in the $(-)$ flat region of the ``inflaton'' $\vp$
(``early'' universe) the WDW eq.\rf{WDW-eq-final} reduces to the quantum
mechanical Schr{\"o}dinger equation:
\br
\Bigl\lb \h \frac{\pa^2}{\pa {\wti a}^2} + 
\cW_{(-)} ({\wti a},p_\vp,p_{\wti u})\Bigr\rb 
\Psi ({\wti a},p_\vp,p_{\wti u}) = 0 \; ,
\lab{WDW-minus} \\
\cW_{(-)} ({\wti a},p_\vp,p_{\wti u})= 
\frac{3M_1^2}{64 (\chi_2 M_2 -2M_0)}\, {\wti a}^2 + 
\frac{8 p^2_\vp}{3}\, {\wti a}^{-2} 
\nonu \\
+ p_{\wti u} \sqrt{\frac{M_1}{\chi_2 M_2 - 2M_0}}
+ O\Bigl(\frac{p^2_{\wti u}}{{\wti a}^2}\Bigr) \; .
\lab{W-minus}
\er
In \rf{WDW-minus}-\rf{W-minus} the canonical ``darkon'' momentum (times a
constant) plays the role of energy eigenvalue 
$E \equiv p_{\wti u} \sqrt{\frac{M_1}{\chi_2 M_2 - 2M_0)}}$, meaning that the
``darkon'' field ${\wti u}$ plays the role of cosmological ``time'' in the 
``early'' universe.

We can solve explicitly WDW eq.\rf{WDW-minus} for small ``darkon'' momenta 
$p_{\wti u}$ ignoring the last term in \rf{W-minus}:
\br
\Psi ({\wti a},p_\vp,p_{\wti u})= \mathrm{const}\;\,
{\wti a}^{\h (1+\sqrt{1-\g})}\, e^{\frac{i}{2}\b {\wti a}^2}
\nonu \\
\times U\Bigl(\frac{1}{4}(2+\sqrt{1-\g}) - i\frac{E}{2\b},\h (2+\sqrt{1-\g}); 
-i\b {\wti a}^2\Bigr) \; ,
\lab{WDW-sol-minus} \\
\b \equiv \sqrt{\frac{3 M_1^2}{32(\chi_2 M_2 - 2M_0}} \quad ,\quad
\g \equiv \frac{64}{3}p^2_\vp \quad ,\quad 
E \equiv p_{\wti u} \sqrt{\frac{M_1}{\chi_2 M_2 - 2M_0}} \; ,
\lab{WDW-sol-minus-defs}
\er
where $U(\cdot,\cdot;z)$ denotes the confluent hypergeometric function of the second
kind. 
Again as in \rf{no-singularity-plus} the wave function \rf{WDW-sol-minus} vanishes
at ${\wti a}=0$:
\be
\Psi ({\wti a},p_\vp,p_{\wti u}) \simeq 
\mathrm{const}\;\, {\wti a}^{\h (1-\sqrt{1-\g})} 
\;\; \mathrm{for} \;\; {\wti a} \to 0 \; ,
\lab{no-singularity-minus}
\ee

In the inflationary ``slow-roll'' regime in the ``early'' Universe the
``inflaton'' canonical momentum $p_\vp$ is very small. Thus, ignoring also the
second term in $\cW_{(-)}$ \rf{W-minus} and Fourier-transforming \rf{WDW-sol-minus}
w.r.t. canonical ``darkon'' momentum $p_{\wti u}$ with $E$ as in 
\rf{WDW-sol-minus-defs}:
\be
\Psi ({\wti a},\t) = \int^{\infty}_{-\infty} \frac{dE}{2\pi} 
\Psi ({\wti a},p_\vp\!=\!0,p_{\wti u})\,e^{-iE\t} \quad ,\;\;
E \equiv p_{\wti u} \sqrt{\frac{M_1}{\chi_2 M_2 - 2M_0}} \; ,
\lab{WDW-sol-minus-tau}
\ee
\textsl{i.e.}, $\t \sim {\wti u}$ being the ``cosmological'' time, the WDW
equation \rf{WDW-minus}-\rf{W-minus} acquires the form of a time-dependent 
Schr{\"o}dinger equation for the inverted harmonic oscillator:
\be
i\partder{}{\t} \Psi ({\wti a},\t) = \Bigl\lb - \h \frac{\pa^2}{\pa {\wti a}^2} 
- \om^2 {\wti a}^2 \Bigr\rb \Psi ({\wti a},\t)
\lab{inverted-oscillator}
\ee
with a negative ``frequency'' squared: 
\be
- \om^2 \equiv - \frac{3M_1^2}{64 (\chi_2 M_2 -2M_0)} \equiv 
- \frac{3}{16} U_{(-)} \; ,
\lab{frequency-inverted}
\ee
where $U_{(-)}$ \rf{U-minus} is the vacuum energy density of the inflationary
epoch.

The solution of Eq.\rf{inverted-oscillator} in the form of a normalized 
(on the semiaxis ${\wti a} \in (0,\infty)$) wave packet has already been 
found in \ct{guth-pi}:
\br
\Psi ({\wti a}, \t) = \Bigl(\frac{2\om}{\pi} \sin(2b)\Bigr)^{1/4}
\bigl(\cos(b-i\om\t)\bigr)^{-1/2}
\nonu \\
\times \exp\{-\h {\wti a}^2 \om \tan(b-i\om\t)\} \; ,
\lab{WDW-sol}
\er
where the parameter $b$ describes the width of the wave packet. 
Calculating the average value of the FLRW scale factor 
$a = \frac{\sqrt{3}}{4} {\wti a}^{2/3}$ (cf. \rf{a-tilde}) we obtain:
\be
\llangle {\wti a} \rrangle \equiv \int_0^{\infty} d{\wti a}\, {\wti a}
|\Psi ({\wti a}, \t)|^2 
= \Bigl\lb \frac{\cos(2b) + \cosh(2\om \t)}{\pi \om \sin(2b)}\Bigr\rb^{1/2} \; .
\lab{a-average}
\ee
Thus, the quantum average of the FLRW scale factor does not exhibit any
singularity ($\llangle {\wti a} \rrangle \to 0$) at any ``time'' $\t$.

\section{Conclusions}
\label{conclude}

Employing non-Riemannian spacetime volume-forms (non-Riemannian volume elements) 
in generalized gravity-matter theories allows for several interesting developments:

\begin{itemize}
\item
Simple unified description of dark energy and dark matter as manifestation
of the dynamics of a single non-canonical scalar field (``darkon'').
\item
Construction of a new class of models of gravity interacting with a scalar
``inflaton'' $\vp$, as well as with other phenomenologically relevant matter
including Higgs-like scalar $\s$, which produce an effective full scalar potential 
of $\vp,\s$ with few remarkable properties.
%
\item
The ``inflaton'' effective potential (at fixed $\s$) possesses two infinitely large
flat regions with vastly different energy scales for large negative and
large positive values of $\vp$. 
This allows for a unified description of both ``early'' universe inflation as
well as of present ``dark energy''-dominated epoch in universe's evolution. 
\item
In the ``early'' universe the would-be Higgs field $\s$ remains massless and
decouples from the ``inflaton'' $\vp$. The ``early'' universe evolution is
described entirely in terms of the ``inflaton'' dynamics. 
\item
In the post-inflationary epoch $\vp$ and $\s$ exchange roles. The inflaton $\vp$ 
becomes massless and decoupled, whereas $\s$ becomes a genuine Higgs field
with a dynamically generated electro-weak-type symmetry breaking effective
potential.
%
\item
A natural choice for the parameters involved conforms to quintessential
cosmological and electro-weak phenomenologies.
\item
Minisuperspace Wheeler-De Witt quantization reveals the role of the
``darkon'' scalar field as cosmological ``time'' in the ``early'' Universe.
The quantum average of the FLRW scale factor does not exhibit any
singularity in its ``time'' evolution.
\end{itemize}

Let us also note that applying the non-Riemannian volume-form formalism to 
minimal $N=1$ supergravity we arrived at a novel mechanism for the 
supersymmetric Brout-Englert-Higgs effect, namely, 
the appearance of a dynamically generated cosmological constant triggering 
spontaneous supersymmetry breaking and mass generation for the gravitino 
\ct{susyssb-1,susyssb-2}. 
Applying the same non-Riemannian volume-form formalism to anti-de Sitter 
supergravity produces simultaneously a very large physical gravitino mass 
and a very small {\em positive} observable cosmological constant 
\ct{susyssb-1,susyssb-2} in accordance with modern cosmological scenarios 
for slowly expanding universe of the present epoch 
\ct{dark-energy-observ-1}-\ct{dark-energy-observ-7}.

As a final comment let us mention some further extensions of the method of
non-Riemannian volume elements -- gravity models with dynamical spacetime
\ct{eduardo-dynamical-time} further developed into models of interacting 
diffusive unified dark energy and dark matter (see \ct{benisty-eduardo} 
and references therein).
%

\begin{acknowledgement}
E.G., E.N. and S.P. gratefully acknowledge support of our collaboration through 
the academic exchange agreement between the Ben-Gurion University in Beer-Sheva,
Israel, and the Bulgarian Academy of Sciences. 
E.N. and E.G. have received partial support from European COST actions
MP-1405 and CA-16104, and from CA-15117 and CA-16104, respectively.
E.N. and S.P. are also thankful to Bulgarian National Science Fund for
support via research grant DN-18/1. 
\end{acknowledgement}
%

%
%

\end{document}